\documentclass[twocolumn,showpacs,preprintnumbers,amsmath,amssymb,superscriptaddress]{revtex4-1}

\usepackage{graphicx}
\usepackage{epsfig}
\usepackage{dcolumn}
\usepackage{bm}

\begin{document}

\title{Neutron capture cross section of unstable $^{63}$Ni: implications for stellar nucleosynthesis}
\author{C. Lederer}
\affiliation{University of Vienna, Faculty of Physics, Vienna, Austria}
\affiliation{Johann-Wolfgang-Goethe Universit\"{a}t, Frankfurt, Germany}%
\author{C. Massimi}\affiliation{Dipartimento di Fisica, Universit\`{a} di Bologna, and Sezione INFN di Bologna, Italy}
\author{S.~Altstadt}\affiliation{Johann-Wolfgang-Goethe Universit\"{a}t, Frankfurt, Germany}%
\author{J.~Andrzejewski}\affiliation{Uniwersytet \L\'{o}dzki, Lodz, Poland}%
\author{L.~Audouin}\affiliation{Centre National de la Recherche Scientifique/IN2P3 - IPN, Orsay, France}%
\author{M.~Barbagallo}\affiliation{Istituto Nazionale di Fisica Nucleare, Bari, Italy}%
\author{V.~B\'{e}cares}\affiliation{Centro de Investigaciones Energeticas Medioambientales y Tecnol\'{o}gicas (CIEMAT), Madrid, Spain}%
\author{F.~Be\v{c}v\'{a}\v{r}}\affiliation{Charles University, Prague, Czech Republic}%
\author{F.~Belloni}\affiliation{Commissariat \`{a} l'\'{E}nergie Atomique (CEA) Saclay - Irfu, Gif-sur-Yvette, France}%
\author{E.~Berthoumieux}\affiliation{Commissariat \`{a} l'\'{E}nergie Atomique (CEA) Saclay - Irfu, Gif-sur-Yvette, France}%
\affiliation{European Organization for Nuclear Research (CERN), Geneva, Switzerland}%
% ...
%

\date{\today}

\begin{abstract}
The $^{63}$Ni($n, \gamma$) cross section has been measured 
for the first time at the neutron time-of-flight facility n\_TOF at CERN from thermal neutron energies up to 200 keV. 
In total, capture kernels of  12 (new) resonances were determined. Maxwellian 
Averaged Cross Sections were calculated for thermal energies
from kT = 5 keV to 100~keV with uncertainties around 20\%. 
Stellar model calculations for a 25 M$_\odot$ star show
that the new data have a significant effect on the $s$-process
production of $^{63}$Cu, $^{64}$Ni, and $^{64}$Zn in massive 
stars, allowing stronger constraints on the Cu yields from explosive nucleosynthesis in the subsequent supernova.
\end{abstract}


\maketitle
The weak component of the astrophysical $s$~process  observed in the Solar System abundance distribution
includes the $s$-process species between Fe and Sr ($60 < A < 90$) \cite{KAEPP89}.
Most of them are generated in massive stars during convective core He-burning
and convective shell C-burning via the activation of the neutron source reaction
$^{22}$Ne($\alpha$,n)$^{25}$Mg \cite{PET68,COUCH74,LAMB77,RAIT91a,RAIT91b}. 
The long-lived radioisotope isotope $^{63}$Ni ($ t _{1/2}$=$101.2\pm1.5$~ yr  
\cite{Colle2008})  is located along the neutron capture path, and in typical weak
$s$-process conditions it may become a branching point, when the neutron capture timescale is comparable with
its stellar $\beta$-decay rate.\\
In particular, at the end of He core burning
the neutron source $^{22}$Ne($\alpha$,n)$^{25}$Mg is activated at
temperatures around 0.3~GK (GK = 10$^{9}$ K), corresponding 
to a Maxwellian neutron energy distribution for a thermal energy of
$kT=26$~keV. At this stage, neutron densities are too weak for a subsequent neutron capture on  $^{63}$Ni (with central peak neutron density in the order of
10$^{7}$ cm$^{-3}$, e.g. \cite{RAIT91a,RAIT91b}) and more than 90\% of the $^{63}$Ni produced 
decays to $^{63}$Cu. However, the $s$-process material is partly reprocessed 
during C shell burning, where the $^{22}$Ne($\alpha, n$)$^{25}$Mg neutron 
source is reactivated at much higher temperatures of about 1~GK, corresponding 
to a thermal energy of $kT=90$~keV. \\
During this second stage, 
neutron densities are orders of magnitudes higher, reaching a maximum of 
10$^{11-12}$ cm$^{-3}$  \cite{PIG10}.  At the $^{63}$Ni branching point, the high neutron densities favor 
neutron capture producing $^{64}$Ni, bypassing the 
production of $^{63}$Cu despite of the strong temperature dependence of the $^{63}$Ni $\beta$-decay rate  
(at C shell temperatures the half-life of $^{63}$Ni decreases to few years \cite{LANG00}).  
In these conditions, the amount of $^{63}$Cu generated during
He core burning is partially depleted in the C shell, but the final $^{63}$Cu abundance will increase thanks to the later radiogenic
decay of the $^{63}$Ni accumulated in the C shell burning phase \cite{PIG10}. In Fig. 1, we show the neutron capture path in the Ni-Cu-Zn region
during He core and at high neutron density during C shell burning.
\begin{figure}[!htb]
\includegraphics[width=5. cm]{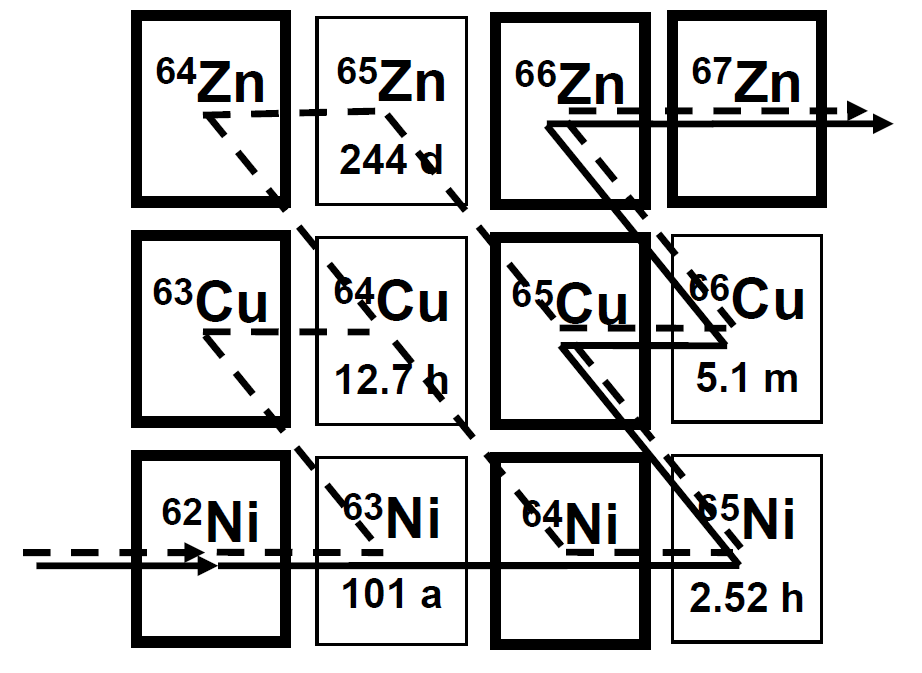}
\caption{ The $s$-process reaction path in the Ni-Cu-Zn region during He core 
burning (dashed lines) and C shell burning (solid lines). \label{path}}
\end{figure}
Up to now the stellar cross section of $^{63}$Ni($n,\gamma$)$^{64}$Ni relied on 
calculations or extrapolations of experimental values at thermal neutron 
energies (0.025~eV) \cite{BARN71,MICH74,Hard92}. Theoretical predictions
for the Maxwellian Averaged Cross Section (MACS) at $kT=30$~keV are ranging 
from 24 to 54 mb \cite{WOO78,HAR81,RAU00,GOR02,GOR05}. 
The currently recommended value quoted by the compilation KADoNiS \cite{kadonis}
is $31\pm6$~mb. Because such calculations are vulnerable to large systematic uncertainties, measurements 
have been attempted at Los Alamos National Laboratory \cite{dance} and
at CERN. In this letter we report on the first experimental results for the 
$^{63}$Ni cross section at stellar energies obtained at the n\_TOF facility 
at CERN. \\
The measurement was performed at the neutron time of flight facility 
n\_TOF located at CERN. Neutrons are produced via spallation reactions of 
20 GeV/c protons from the Proton Synchrotron with a massive Pb target. 
With the high intensity of the pulsed proton beam, the repetition rate of 
0.4 Hz, a short proton pulse width of 6 ns, and a neutron flight path of 185 m, 
the n\_TOF facility is unique for the combination of a high neutron energy 
resolution and a high instantaneous neutron flux. A more detailed 
description of the facility can be found in Ref. \cite{guerrero2012} and references therein.  \\
The $^{63}$Ni  sample was produced by irradiating highly enriched 
$^{62}$Ni in a thermal reactor \cite{Hard92,Muthig1984,Trautmannsheimer1992}. 
Since the irradiation took place more than 20 years ago, the original $^{63}$Ni
fraction had partially decayed to $^{63}$Cu. To avoid background due to 
$^{63}$Cu($n, \gamma$)$^{64}$Cu reactions, this $^{63}$Cu impurity has been 
chemically separated prior to the ($n, \gamma$) measurement. The 
originally metallic target was dissolved in concentrated nitric acid and the 
copper fraction was precipitated as CuS using gaseous H$_{2}$S.
The remaining solution was treated with NaOH to precipitate Ni(OH)$_{2}$,
which was calcinated at 800~$^{\circ}$C to form NiO. By means of mass 
spectrometry, the $^{63}$Ni/$^{62}$Ni ratio in the sample was determined 
to $0.123\pm0.001$\%, and the contribution from other Ni isotopes was found to 
be $\leq1$\%. In total, 1156 mg NiO powder were encapsulated in a thin-walled cylinder made of PEEK (Polyetheretherketone, 
net mass 180 mg) to produce a sample 20 mm in diameter and 2.2 mm in
thickness. \\
The neutron capture yield was measured as a function of neutron energy 
by detecting the prompt capture $\gamma$ rays with a pair of liquid 
C$_6$D$_6$ scintillation detectors. These detectors are optimized to exhibit 
a very low sensitivity to neutrons, thus minimizing the background produced 
by neutrons scattered on the sample \cite{plag}. The dependence of the detection 
efficiency on $\gamma$-ray energy and the effect of the $\gamma$-ray 
threshold of 250 keV were corrected using the Pulse Height Weighting technique 
\cite{pwht1,pwht2}. By application of a pulse-height dependent weight on the 
deposited $\gamma$ energy the detection efficiency becomes a linear function 
of the excitation energy of the compound nucleus, $\varepsilon\approx k\times E_c$. 
Choosing $k=1$~MeV$^{-1}$, the capture yield can be obtained as
\begin{equation}
 Y=N\frac{C_w}{\Phi E_c}
\end{equation}
where $C_w$ are the weighted, background-subtracted counts, $\Phi$ denotes 
the relative neutron flux, and $N$ a normalization factor for the absolute capture yield.
The normalization factor was determined via the saturated resonance technique 
\cite{Mack1979} in an additional run with a Au sample of the same size as the Ni 
sample. The Au sample was chosen such that the gold 
resonance at 4.9~eV is saturated, which means that all neutrons of that energy are 
absorbed in the sample, thus providing a measure for the absolute neutron flux at 4.9 eV. 
The energy dependence of the neutron flux was measured relative to the standard 
cross sections $^{10}$B(n,$\alpha$) and $^{6}$Li($n,\alpha$) up to 150~keV and $^{235}$U($n,f$) at higher energies.
Because the size of the neutron beam widens slightly with neutron energy, the normalization factor $N$, related to the fraction of the neutron beam intercepted by the sample, 
changes as well. This effect was taken into account by simulations of the neutron beam profile \cite{guerrero2012}. \\
The experimental background was determined in dedicated runs with an empty 
PEEK container, with a $^{62}$Ni sample of the same diameter, and in runs without 
neutron beam. Additionally, the neutron capture yield has been measured with 
a set of neutron filters located about 50 m upstream of the sample. These W, Mo, 
and Al filters are thick enough to exhibit black resonances at certain energies, so 
that all neutrons in these windows are completely removed from the beam and do 
not reach the sample. Accordingly, the level in the corresponding dips in these 
spectra is expected to represent the experimental background. When the run with 
filters was repeated using only the empty container, the same background 
level was observed in the filter dips, thus confirming the background measured
with the empty sample container.
\begin{figure}[!htb]
\includegraphics[width=8 cm]{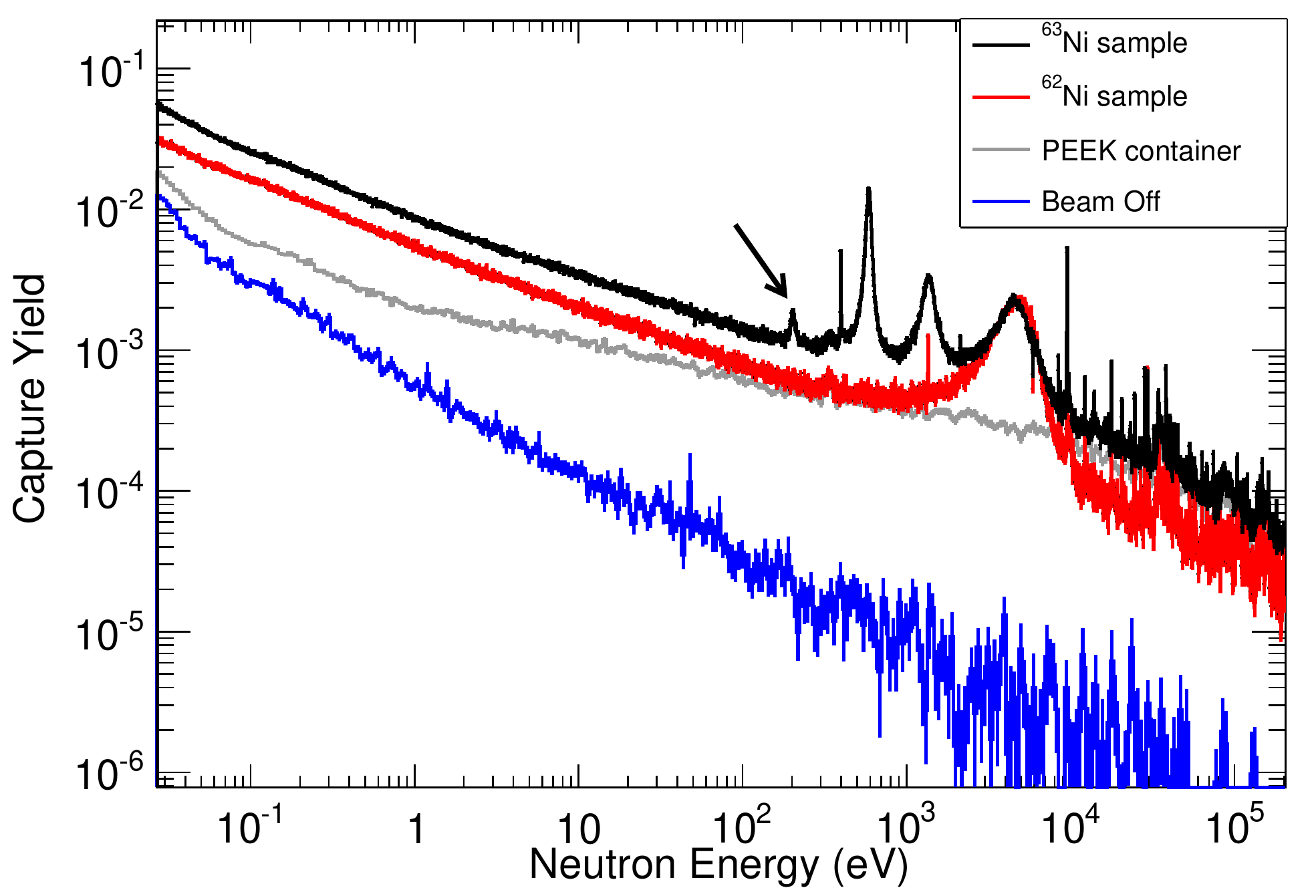}
\caption{(Color online) Capture yield of the $^{63}$Ni sample (black) compared 
to the empty sample holder and to the spectrum obtained with a pure 
$^{62}$Ni sample. The background from the activity of the sample
and from ambient radiation (blue) is almost negligible at keV energies. 
The capture yield of the $^{62}$Ni sample was scaled for the $^{62}$Ni mass present in the
$^{63}$Ni sample. The first resonance at 203 eV (marked by an arrow) is assigned to a small impurity of $^{59}$Ni in the sample.\label{yield}}
\end{figure}
Figure \ref{yield} shows a comparison of the capture yield of the $^{63}$Ni 
sample, the empty PEEK container and the $^{62}$Ni sample. The background 
from the radioactivity of the $^{63}$Ni sample and from ambient radiation, 
which was obtained from a measurement without neutron beam, is given 
as well. Between 100~eV and 2~keV four resonances are visible in 
the spectrum of the $^{63}$Ni sample, which are obviously not correlated 
with the $^{62}$Ni($n,\gamma$) content.  The first of these resonances (marked by an arrow) can be attributed to a resonance in the 
$^{59}$Ni($n,\gamma$) reaction \cite{Mugh06}, 
expected at that neutron energy and compatible with the measured 0.03\% impurity of $^{59}$Ni in our sample.
The other three resonances are clearly attributable to the $^{63}$Ni($n, \gamma$)
channel.  This holds also for several other resonances up to neutron energies of 
55~keV, for which the capture kernels
\begin{equation}
A_\gamma = \frac{1}{2\pi^2\lambdabar^2}\int_{-\infty}^{+\infty}{ \sigma(E) dE}=g_s\frac{\Gamma_n\Gamma_\gamma}{\Gamma_n+\Gamma_\gamma}
\end{equation}
characterizing the strength of the resonance, could be deduced by a resonance shape analysis (RSA) with the R-matrix code SAMMY \cite{sammy}
(Table \ref{tab3}). The capture kernel is determined by the spin statistical factor $g_s$, the neutron width $\Gamma_n$, and the radiative width $\Gamma_\gamma$. 
For two resonances also the orbital angular 
momentum $\ell$, derived from the shape of the resonance, is given.
 The neutron energy interval between 2 and 8~keV is dominated by the strong resonance in 
$^{62}$Ni($n, \gamma$) at 4.6~keV, therefore smaller resonances in $^{63}$Ni($n, \gamma$)
might be invisible due to this background. In summary, 12 levels in $^{64}$Ni were observed for the first time.

\begin{table}[htb]
\caption{\label{tab3}Resonance energies $E_r$ (laboratory energy)
and capture kernels $A_\gamma$ for the $^{63}$Ni($n, \gamma$) reaction. For resonances marked with an asterisk
the orbital angular momemtum $\ell=0$ could be deduced from the resonance shape. 
}
\begin{ruledtabular}
\begin{tabular}{cc|cc}

$E_r$ (eV) &  $A_\gamma$ (meV) &  $E_r$  & $A_\gamma$ (meV) \\ \hline
$397.96\pm0.04 $&  $  5.7\pm0.4 $& $9776\pm3     $  & $     100\pm10$\\
$587.25\pm0.09^{*}$&  $  340\pm20$ &$13984\pm3   $  & $  131\pm45$\\
$1366\pm1^{*} $& $   810\pm40$ & $17127\pm4    $ &  $       108\pm59$\\
$8634\pm2     $  & $     45\pm9$ & $19561\pm6    $  & $    130\pm20$\\
$8981\pm3     $  & $     50\pm10$  & $32330\pm10   $  &  $  500\pm200$\\
$9154\pm4     $  & $     43\pm9$ & $54750\pm30  $   & $  700\pm200$\\ 
\end{tabular}\end{ruledtabular}
\end{table}

As a consequence of the small sample mass, the signal to background ratio 
starts to deteriorate already above 10~keV. Accordingly, it is increasingly 
difficult to identify resonances with confidence at higher energies. 
Thus, MACSs were calculated using resonance 
parameters only below 10~keV, whereas averaged cross section data have 
been determined from 10~keV to 200~keV.  These data were obtained by 
subtraction of the yield measured with the $^{62}$Ni sample after it 
had been properly scaled for the $^{62}$Ni content of the $^{63}$Ni 
sample. The background due to oxygen is negligibly small because of its
 very small ($n, \gamma$) cross section.

\begin{table}[htb]
\caption{\label{tab2}Maxwellian Averaged Cross Sections (in mb) of $^{63}$Ni($n,\gamma$)
compared to previously recommended values based on theoretical predictions \cite{kadonis}. 
The respective contributions from resonances below 10 keV (RC) are listed separately.
Uncertainties are 1~$\sigma$.}
\begin{ruledtabular}

\begin{tabular}{cccc}
$kT$     & KADoNiS&\multicolumn{2}{c}{This Work}\\
\cline{3-4}
(keV)    &  & RC        &  Total \\
\hline
5       &112      &   $174\pm6$      & ~$224\pm8_\text{stat}\pm45_\text{sys}$ \\
10       &66       &$51\pm2$      & $129.5\pm7.1\pm25.9$  \\
15       &50       & $24\pm1$       & $101.3\pm6.9\pm20.3$ \\
20       &41       &$14\pm1$    & ~$85.5\pm6.4\pm17.1$   \\
25       &35       &~$9.3\pm0.4$   & ~$74.9\pm5.9\pm15.0$ \\
30       &31$\pm$6  &~$6.6\pm0.3$   & ~$66.7\pm5.4\pm13.3$   \\
40       &25       &~$3.8\pm0.2$   & ~~$54.5\pm4.6\pm10.9$ \\
50       &20       &~$2.4\pm0.1$   & $45.6\pm3.9\pm9.1$  \\
60       &17       &~$1.7\pm0.1$   & $38.8\pm3.4\pm7.8$    \\
80       &13    &~$0.97\pm0.05$ &  $29.1\pm2.7\pm5.8$   \\
100       &10    &~$0.63\pm0.03$ &  $22.5\pm2.1\pm4.5$   \\
\end{tabular}
\end{ruledtabular}
\end{table}

The MACSs for thermal energies from 
$kT=5$ to 90~keV are listed in Table \ref{tab2}, together with the theoretical predictions
in the KADoNiS compilation ~\cite{kadonis}. Our results are approximately a factor of 2 higher than the calculated cross section. 
The total systematic uncertainties in our results of 20\% are mainly due to subtraction of the background and
the effect of sample impurities - particularly in the region between 2 and 8~keV, where the 
spectrum is dominated by the 4.6 keV resonance in $^{62}$Ni($n,\gamma$). 
Comparably minor contributions to the systematic uncertainty are caused by the neutron flux (3-5\%), the Pulse Height Weighting technique
(2\%), the flux normalization (1\%), and the $^{63}$Ni/$^{62}$Ni ratio (1.6\%).

\begin{figure}[!htb]
\includegraphics[width=7. cm]{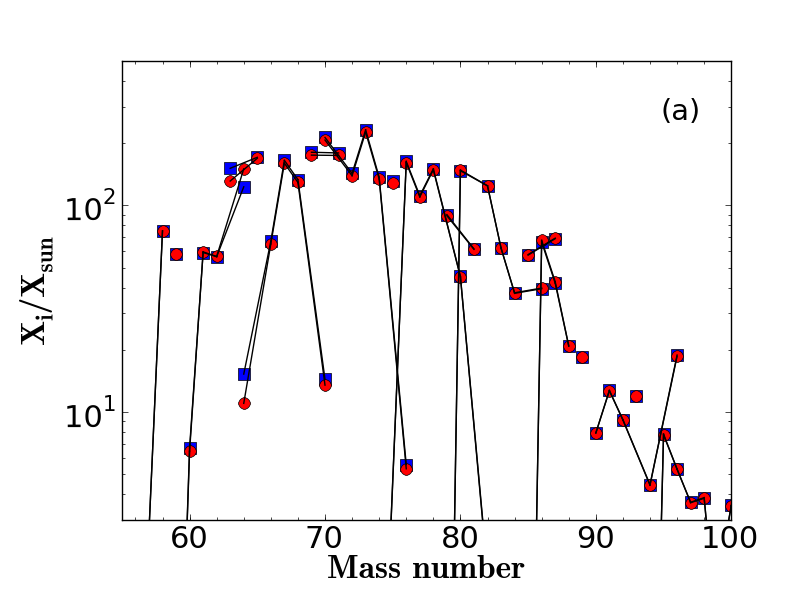}
\includegraphics[width=7. cm]{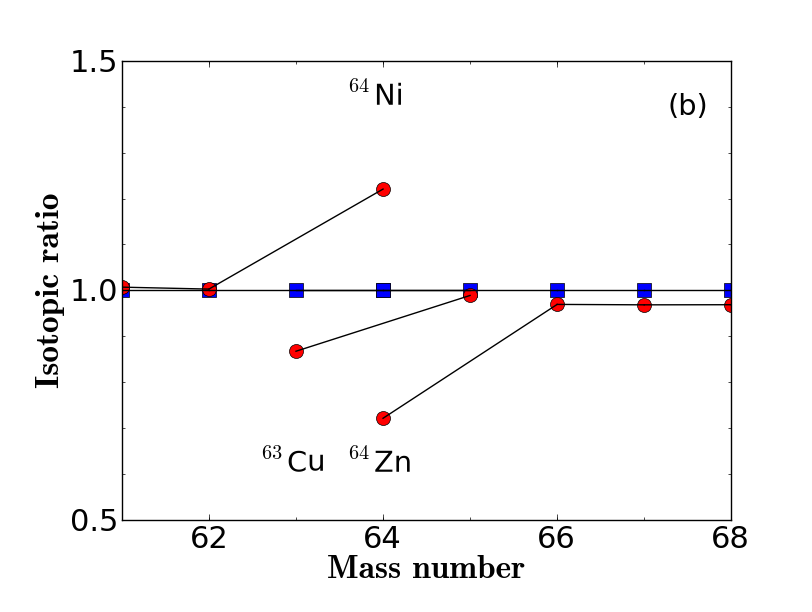}
\caption{\label{calc}(Color online) (a) Final isotopic $s$-process distributions using the new
measured $^{63}$Ni MACS (red circles) and the MACS quoted by KADoNiS (blue squares) \cite{kadonis}. The distribution is 
normalized to solar system abundances. 
(b) Ratio of the two distributions in Panel (a), zoomed in the Ni-Cu-Zn
mass region. Isotopes of the same element are connected by solid lines. }
\end{figure}

The impact of our new results was investigated for the $s$~process in a full stellar model for a 25 M$_\odot$ 
star with an initial metal content Z =0.02. The complete nucleosynthesis was followed with the post-processing NuGrid code MPPNP 
\cite{HER08}.  The stellar rates were obtained by combining the measured $^{63\text{gs}}$Ni$(n,\gamma)^{64}$Ni
cross sections and theoretically predicted 
contributions to the stellar rate due to $^{63}$Ni$^*(n,\gamma)^{64}$Ni reactions as described in Ref. \cite{RAU12}. While the 
contribution of $^{63\text{gs}}$Ni$(n,\gamma)^{64}$Ni reactions to the stellar rate is still around 90\% at He Core 
burning temperatures, it drops to around 40\% at the higher temperature in the C Shell burning phase. Because of the 
larger uncertainties of the $^{63}$Ni$^*(n,\gamma)^{64}$Ni cross sections, the uncertainty of the stellar rate increases with 
temperature. Apart  from the reaction cross sections, the final abundance pattern is also affected by the temperature 
dependence of the radioactive decay rates under stellar conditions. In the investigated mass region the concerned 
rates for the $\beta^-$-decay of $^{63}$Ni and the $\beta^+$/$\beta^-$-decays of $^{64}$Cu have been adopted 
from Ref. \cite{LANG00}. By variation within reasonable limits \cite{PIG10} it was found that the decay rates of both isotopes have a comparably 
small effect on the investigated abundances, because the reaction flow in the $^{63}$Ni branching is governed by 
the neutron density conditions, which lead either to much lower or much higher ($n, \gamma$) rates during core He and shell C burning, respectively.\\
The calculated abundance distribution from Fe to Zr shown in Fig. \ref{calc} represents the $s$~abundances after core He 
and shell C burning, i.e. prior to the supernova explosion, at a point where the nucleosynthesis yields are well 
characterized by the model \cite{PIG10}. The distribution is compared in Fig. \ref{calc} to the one obtained with the 
neutron capture rates of the KADoNiS evaluation \cite{kadonis}. The ratio of the two distributions in 
the lower panel of Fig. \ref{calc} shows that the new $^{63}$Ni cross section affects only a few isotopes between Ni and
Zn. An enhancement of about 20\% is found for $^{64}$Ni, while $^{63}$Cu is depleted by about 15\%. As
the $^{65}$Cu yields remain essentially unchanged, the isotopic ratio $^{63}$Cu/$^{65}$Cu is correspondingly reduced 
at the end of shell C burning. $^{64}$Zn is depleted as well (by about 30\%), because  $^{63}$Cu and  $^{64}$Zn are populated by the nucleosynthesis channel following the 
$\beta^-$ branch $^{62}$Ni($n, \gamma$)$^{63}$Ni($\beta^-$)$^{63}$Cu($n, \gamma$)$^{64}$Cu($\beta^-$)$^{64}$Zn. 
However, the $s$-process contribution to $^{64}$Zn remains marginal as this isotope results predominantly from 
later explosive nucleosynthesis during Core Collapse Supernovae \cite{HWF06}. Also the propagation effect of the new MACS of $^{63}$Ni 
on heavier $s$-process species is rather small, of the order of a few percent. \\
Although the $s$~process component at the end of convective shell C burning is well defined
by these calculations, the abundances in the Ni-Cu-Zn region may be affected by following burning 
stages (for instance the possible merging of shells \cite{RAU02}) and by the subsequent supernova explosion before enriching the interstellar medium. Given the 
complexity of this scenario, the final abundances are yet subject to considerable uncertainty as emphasized 
by several sensitivity studies \cite{PIG10,RAU02,THA09}. Nevertheless, the present results represent a fundamental improvement in constraining the 
weak $s$-process component from the convective core He burning and convective C shell burning phases.
A better knowledge of the pre-explosive weak $s$-process component will allow to better define also the following explosive contribution to the copper inventory, once robust 
theoretical predictions are compared with spectroscopic observations. Another relevant observational constraint is given by the copper isotopic ratio in the Solar System, where $s$~process in massive
stars provide the dominant contribution (\cite{PIG10}, and references therein).\\
In summary, we measured the energy-dependent $^{63}$Ni($n,\gamma$) cross section at the n\_TOF facility providing the first experimental results for MACSs at stellar neutron energies. 
The MACSs ranging from $kT=5$ to 90~keV exhibit total uncertainties of $20-22$\% and are about a factor of 2 higher than the theoretical prediction of the KADoNiS compilation. 
Our results improve one of the main nuclear uncertainties affecting theoretical predictions for the abundances of $^{63}$Cu, $^{64}$Ni and $^{64}$Zn in $s$-process rich ejecta of core collapse supernovae.
Furthermore, these results are a fundamental step to constrain the contribution from explosive nucleosynthesis to these species.\\

The authors would like to thank H.~Danninger and C.~Gierl of the Technical University of Vienna for their help preparing the $^{62}$Ni sample. 
This work was partly supported by the Austrian Science Fund (FWF), projects P20434 and I428. MP also acknowledges the support from the Ambizione grant of the Swiss NSF,
 the  NSF grants PHY 02-16783 and PHY 09-22648 (Joint Institute for Nuclear Astrophysics, JINA),  EU grant MIRG-CT-2006-046520, and EuroGenesis.
TR acknowledges support from EuroGenesis, the FP7 ENSAR/THEXO project and by a "Distinguished Guest Scientist Fellowship" from the Hungarian Academy of Sciences.

\end{document}